\documentclass[twoside]{article}
\usepackage{qic,epsfig}

\usepackage{amsmath,bbm,graphicx}

\newtheorem{observation}[theorem]{Observation}

\newcommand{\ket}[1]{\mbox{$| #1 \rangle$}}

\newcommand{\square}{\hfill\fbox\\\medskip }

\def\ZZ{\mathbbm{Z}}
\def\RR{\mathbbm{R}}
\def\CC{\mathbbm{C}}

\def\NN{\mathbbm{N}}

\def\H{\mathcal{H}}
\def\Id{\mathbbm{1}}

\DeclareMathOperator{\tr}{tr}

\begin{document}

\setlength{\textheight}{8.0truein}    

\runninghead{Quantum Margulis expanders}
            {D.\ Gross and J.\ Eisert}

\normalsize\textlineskip
\thispagestyle{empty}
\setcounter{page}{1}

\copyrightheading{0}{0}{2003}{000--000}

\vspace*{0.88truein}

\alphfootnote

\fpage{1}

\centerline{\bf
Quantum Margulis expanders}
\vspace*{0.37truein}
\centerline{\footnotesize
D.\ Gross and J.\ Eisert
}
\vspace*{0.015truein}
\centerline{\footnotesize\it 
	Institute for Mathematical Sciences, 
	Imperial College London,
}
\baselineskip=10pt
\centerline{\footnotesize\it 
	Prince's Gate, London, SW7 2BW, UK
}
\baselineskip=10pt
\centerline{\footnotesize
	and
}
\baselineskip=10pt
\centerline{\footnotesize\it 
	QOLS, Blackett Laboratory, Imperial 
	College London,
}
\baselineskip=10pt
\centerline{\footnotesize\it 
	Prince Consort Road, London SW7 2BW,
	UK
}
\vspace*{10pt}

\vspace*{0.225truein}
\publisher{(received date)}{(revised date)}

\vspace*{0.21truein}

\abstracts{
We present a simple way to quantize the well-known Margulis expander
map. The result is a quantum expander which acts on discrete Wigner
functions in the same way the classical Margulis expander acts on
probability distributions. The quantum version shares all essential
properties of the classical counterpart, e.g., it has the same degree
and spectrum. Unlike previous constructions of quantum expanders, our
method does not rely on non-Abelian harmonic analysis. Analogues for
continuous variable systems are mentioned. 
Indeed, the construction seems one of the few instances where applications
based on discrete and continuous phase space methods can be developed 
in complete analogy.
}{}{}

\vspace*{10pt}

\vspace*{3pt}
\communicate{to be filled by the Editorial}

\vspace*{1pt}\textlineskip    

%
%
%
%
%
%
%
%

Motivated by the prominent role expander graphs play in theoretical
computer science \cite{survey}, quantum expanders have recently
received a great deal of attention
\cite{hastings1,hastings2,ben1,ben2,harrow,ambainis}.  In this
work, we report an observation which allows for the simple explicit
construction of such quantum expanders. The method relies heavily on
quantum phase space techniques: Once familiar with this techniques,
the result is an almost trivial corollary of the analogous classical
statement. We further discuss continuous analogues of quantum
expanders, where again, phase space methods render this an obvious
generalization. 
Hence, the present work can equally be regarded as the presentation
of a simple quantum expander, as as a short exposition of the
strengths of the phase space formalism as such.

\section{Preliminaries}

\subsection{Expanders}

Expander graphs turn up in various areas of combinatorics and computer
science (for all claims made in this section, the reader is referred to
the excellent survey article Ref.\ \cite{survey}). They often come
into play when one is concerned with a property which ``typically''
holds, but defies systematic understanding. A simple example is given
by classical error correction codes. One can show that a randomly
chosen code is extremely likely to have favorable properties, but it
seems very difficult to come up with a deterministic construction of
codes which are ``as good as random''.  Expander graphs can be
explicitly constructed, but capture some aspects of generic graphs. It
turns out that this property can be used to de-randomize, e.g., the
construction of codes or certain probabilistic algorithms.

The formal definition is straightforward. Consider a graph $G$ with
$N$ vertices $V$, each having $D$ neighbors (we allow for multiple
links and self-links). There is an obvious way to define a random walk
on the graph: At each time step, a particle initially located on a
vertex $v$ will be moved to one of the $D$ neighbors of $v$ with equal
probability. The resulting Markov process is described by an $N\times
N$ doubly stochastic matrix $A$. The largest eigenvalue of $A$ is
$\lambda_1=1$, corresponding to the ``totally mixed'' eigenvector
$ (1,\dots,1)/N$. 
Let $\lambda$ be the absolute value of the second
largest (by absolute value) eigenvalue. A small value of $\lambda$
means that the Markov process is strongly mixing, i.e., converges
rapidly to the totally mixed state. We call $G$ an
\emph{$(N,D,\lambda)$ expander} if it is described by these
parameters. The goal is to find families of expander graphs with
arbitrarily many vertices $N$, but constant (and small) degree $D$ and
$\lambda$.

While the notion of an expander \emph{graph} seems hard to quantize
(see, however, Ref.\ \cite{hastings2}), it makes sense to look for
quantum analogues of strongly mixing Markov processes with low degree.
Indeed, we call a completely positive map $\Lambda$ a
\emph{$(N,D,\lambda)$-quantum expander} if $\Lambda$ can be expressed
in terms of $D$ Kraus operators acting on $\mathcal{B}(\CC^N)$ and the
absolute value of its second largest singular value is bounded from
above by $\lambda$ (here, $\mathcal{B}(\H)$ denotes the space of linear
operators acting on a linear space $\H$). Once more: The intuition is
to have a quantum channel which can be written using few Kraus
operators, but which rapidly sends any input to the completely mixed
state under repeated invocation \cite{Mixed}.

Quantum expanders have been introduced independently in Ref.\
\cite{hastings1} in the context of Hamiltonian complexity
for the purpose of constructing states of 
spin-chains with certain extremal entanglement and 
correlation properties, and in
Ref.\ \cite{ben1}, where the problem was approached from a computer
science perspective. Very recently, randomized \cite{hastings2} and
explicit \cite{hastings1,ben1,ben2,harrow} constructions of expanders
have appeared in the literature. The basic idea of a quantum expander 
is implicit in earlier work \cite{ambainis}.

\begin{figure}
	\centerline{
	\begin{tabular}{cccc}
  \centering
	0\hspace{-2em}
  \includegraphics[scale=.30]{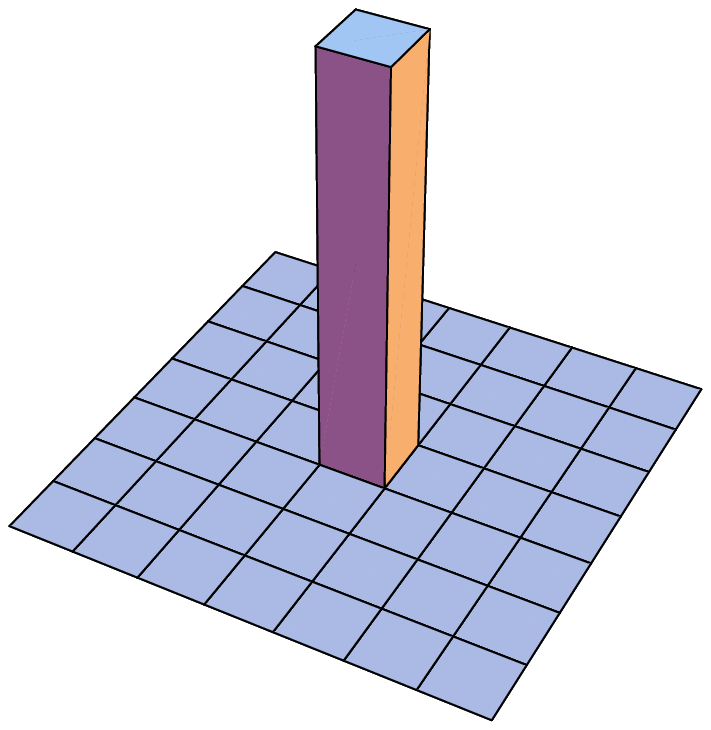}
  \quad
	1\hspace{-2em}
  \includegraphics[scale=.30]{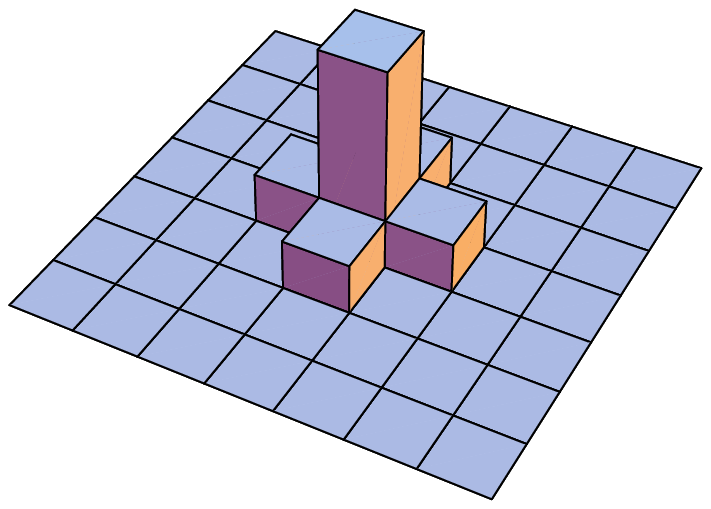}
  \quad
	2\hspace{-2em}
  \includegraphics[scale=.30]{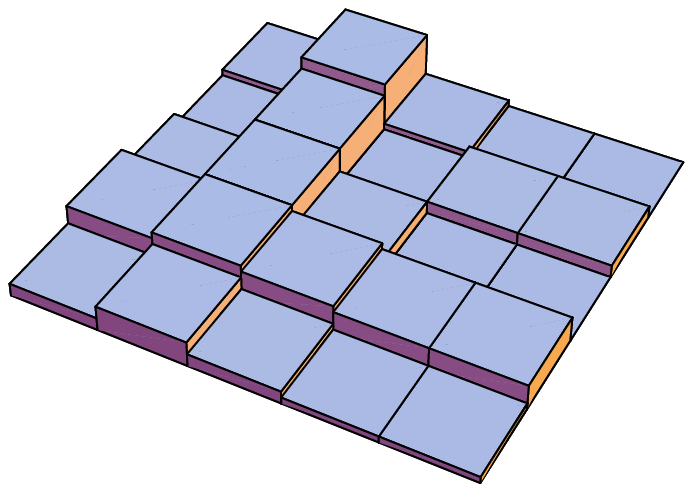}
  \quad
	3\hspace{-2em}
  \includegraphics[scale=.30]{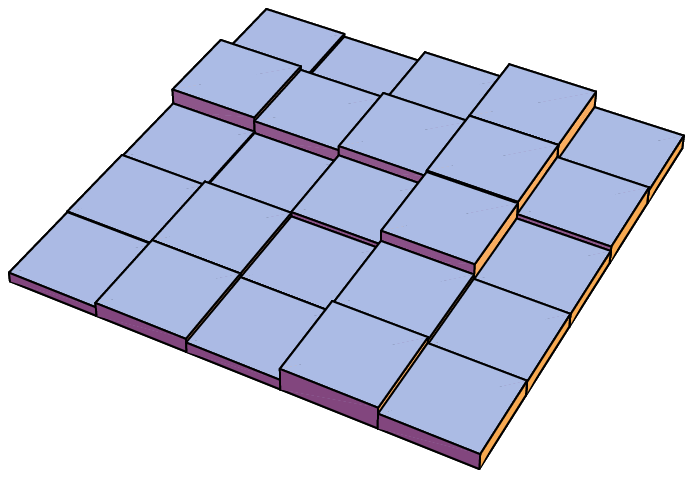}
	\end{tabular}
	}
	\fcaption{
		\label{fig:blur}
		The phase space distributions resulting from three applications of
		the Margulis expander acting on a configuration initially
		concentrated at the origin of a $7\times 7$ lattice. The starting
		distribution can be interpreted either as a classical particle
		with a well-defined position on a two-dimensional lattice, or as
		the quantum phase space operator $A(0,0)$ (see text for
		definition).
	}
\end{figure}

\subsection{Margulis expander}

Margulis provided the first explicit construction of a family of
expander graphs \cite{margulis}. 
Their expansion properties can be verified by
elementary (if tedious) means \cite{survey}. 

The vertices of Margulis' graph are given by the points of a $N\times
N$-lattice.\footnote{
	Note a slight inconsistency in our notation: the number of vertices
	in this case is $N^2$, not $N$.
} We label the axes of the lattice by the elements of
$\ZZ_N=\{0,\dots,N-1\}$. Now consider the four affine transformations
on $\ZZ_N^2$ given by
\begin{eqnarray}
	\label{eqn:margulis}
	T_1&:&\>v \mapsto S_1\,v,\\
	T_2&:&\>v \mapsto S_1\,v+(1,0)^T,\nonumber\\
	T_3&:&\>v \mapsto S_2\,v,\nonumber\\
	T_4&:&\>v \mapsto S_2\,v-(0,1)^T,\nonumber
\end{eqnarray}
where
\begin{equation*}
	S_1= 
	\left[
	\begin{array}{cc}
		1&2\\
		0&1
	\end{array}
	\right],
	\quad
	S_2=
	\left[
	\begin{array}{cc}
		1&0\\
		2&1
	\end{array}
	\right].
\end{equation*}
All operations are modulo $N$. Let ${\cal S}$ be the set of these four
operations, together with their inverses. In Margulis' construction,
two vertices are considered adjacent if and only if they can be mapped
onto each other by an operation in ${\cal S}$.

One finds that $\lambda$ is bounded above by $\sqrt 2\, 5/8$,
independent of $N$ \cite{gabber,survey}. An instance of a 
random walk
on the Margulis graph is visualized in Fig.\ \ref{fig:blur}.

\subsection{Discrete phase space methods}
\label{sec:phaseSpace}

The present section can be approached from a purely mathematical, or
from a physically-oriented point of view. To make the argument more
accessible, we will briefly outline both approaches before going into
details.

\emph{Mathematically}, one starts by noting that the operations $T_i$
used in the construction of the Margulis graph (Eq.\
(\ref{eqn:margulis})) are affine transformations on $\ZZ_N^2$, where
in addition the linear part $S_i$ has unit-determinant. The functions
of this kind form a finite group, which we will refer to as $G_N$. Two
facts will be established below. Firstly,
\begin{itemize}
	\item
	there is a (projective) unitary representation 
	\begin{equation*}
		T\mapsto U_T
	\end{equation*}
	of $G_N$ on $\CC^N$.
\end{itemize}
This representation facilitates the quantization of the expander.
Indeed, the quantum Margulis expander will be defined as the c.p. map
$\Lambda_N$, which applies one of the unitaries $U_T, T\in{\mathcal
S}$ at random. To prove that this construction defines an expander,
we will need a second fact:
\begin{itemize}
	\item
	Let $N$ be odd.  There are $N^2$ hermitian operators
	$A(a)\in\mathcal{B}(\CC^N)$, labeled
	by the points $a\in\ZZ_N^2$, such that
	\begin{enumerate}
		\item
			The operators form an orthonormal basis with respect to the
			Hilbert-Schmidt inner product:
			\begin{equation}\label{eqn:aBasis}
				\frac1N \tr\big(A(v)\,A(w)\big) = \delta_{v,w}.
			\end{equation}

		\item
			The basis thus defined is compatible with the unitary
			representation of $G_N$ in that
			\begin{equation}\label{eqn:cov}
				U_T\,A(v)\,U_T^\dagger = A(T(v)),
			\end{equation}
			for $T\in G_N$, $v\in\ZZ_N^2$.
	\end{enumerate}
\end{itemize}
In order to analyze the action of $\Lambda_N$ on a density operator
$\rho$, we will use Eq.\ (\ref{eqn:aBasis}) to expand $\rho$ in terms
of the $A(a)$'s and then Eq.\ (\ref{eqn:cov}) to reduce the problem to
the classical case (see Section \ref{sec:qMargulis}).

\emph{Physically} speaking, we will employ a phase space description
of the quantum system.  Recall that the term \emph{phase space}
originates in classical mechanics. Here, the state of a single
particle in one spatial dimension is completely specified by two real
parameters: its position and its momentum. The two-dimensional real
vector space spanned by the position and the momentum axes is referred
to as the particle's phase space. Likewise, the state of a single
quantum system can be specified by a quasi-probability distribution on
phase space, namely the particle's Wigner function. The Wigner
function shares many properties of classical probability
distributions, except for the fact that it can take negative values
(see Ref.\ \cite{hudson,poswig,poswigShort} for an analysis of quantum
states which exhibit only positive values).  

In the context of continuous-variable systems, affine
volume-preserving transformations of the phase space are known as
\emph{canonical transformations}.  Let $\rho$ be a density matrix and
denote by $W_\rho(v)$ the associated Wigner function.  It is
well-known \cite{Braun} that for every canonical transformation $T$,
there is a unitary operator $U_T$ which implements the map $T$ in the
sense that
\begin{equation*}
	W_{U_T\rho U_T^\dagger}(v)=W_\rho(T(v)).
\end{equation*}
As detailed below, a similar relation holds for finite-dimensional
quantum systems, associated with discrete phase spaces. Indeed, the
Wigner function of a density operator $\rho$ turns out to be nothing
but the collection of  expansion coefficients of $\rho$ with respect
to the basis given in Eq.\ (\ref{eqn:aBasis}); canonical
transformations are elements of $G_N$; and the correspondence $T\mapsto
U_T$ is just the representation mentioned in the first paragraph of
this section.

So in this physical language, the basic realization is that the
building blocks of the Margulis scheme (Eq.\ (\ref{eqn:margulis}))
are canonical transformations of a discrete phase space.

To make all this more precise, 
let $N$ be odd\footnote{
	We restrict attention to odd dimensions, as the theory of discrete
	Wigner functions is much more well-behaved in
	this case.},
$\H=\CC^N$ and assume
that some basis $\{\ket 0, \dots, \ket {N-1}\}$ in $\H$ has been chosen.
Let $\omega=e^{\frac{2\pi}N i}$ 
be an $N$th root of unity. We define the \emph{shift} and \emph{boost}
operators as the generalizations of the $X$ and $Z$ Pauli matrices by
\begin{eqnarray}\label{eqn:shiftClock}
  x(q)\ket k = \ket{k+q}, \quad\quad
	z(p)\ket k = \omega^{p k} \ket k
\end{eqnarray}
(arithmetic is modulo $N$). The \emph{Weyl operators} are 
\begin{equation}\label{eqn:weyl}
	w(p,q)=\omega^{-2^{-1}p q} z(p)x(q),
\end{equation}
where $2^{-1}=(N+1)/2$ is the multiplicative inverse of $2$ modulo
$N$. For vectors $a=(p,q)\in \ZZ_N^2$, we write $w(a)$ for $w(p,q)$.
Let 
\begin{equation}\label{eqn:parity}
	A(0,0): \ket x \mapsto \ket{-x}
\end{equation}
be the \emph{parity operator} and denote by $A(p,q)$ its 
translated version,
\begin{equation}\label{eqn:phasespaceoperators}
	A(p,q)=w(p,q)\,A(0,0)\,w(p,q)^\dagger.
\end{equation}
We will refer to the $A(p,q)$'s as \emph{phase space operators}.  One
can check by direct calculation that Eq.\ (\ref{eqn:aBasis}) holds.
The \emph{Wigner function} of an operator $\rho$ is the collection of
the expansion coefficients of $\rho$ with respect to the basis formed
by the phase space operators.  Formally:
\begin{equation}\label{eqn:wigner}
	W_\rho(p,q)=\frac1N \tr\big(A(p,q)\,\rho\big).
\end{equation}

There are two symmetries associated with a phase space: translations
and volume-preserving linear operations. We shortly look at both in
turn.  Firstly, it is simple to verify that for $a,b\in \ZZ_N^2$
\begin{equation}\label{eqn:shift}
	w(a)\,A(b)\,w(a)^\dagger=A(a+b).
\end{equation}
Hence, Weyl operators implement translations on phase space. Secondly,
let $S$ be a unit-determinant matrix with entries in $\ZZ_N$. It turns
out \cite{marcus,neuhauser,poswig} that there exists a unitary
operator $\mu(S)$ such that, for all
$a\in\ZZ_N^2$ the relation
\begin{equation}\label{eqn:meta}
	\mu(S)\,A(a)\,\mu(S)^\dagger=A(S\,a)
\end{equation}
holds\footnote{
The operator $\mu(S)$ is the \emph{metaplectic representation} of the
symplectic matrix $S$. In quantum information theory, the set $\{
w(a)\,\mu(S)\,:\,a\in \ZZ_N^2, \det (S)=1\}$ is called the
\emph{Clifford group} \cite{gottesman}, which must to be confused with the
Clifford group appearing in the context of Fermions or representation
theory of $SO(n)$.}.

It follows immediately that for every affine transformation $T$ of the
type given in Eq.\ (\ref{eqn:margulis}), there exists a unitary
operator $U_T$ such that
\begin{equation}
	\label{eqn:affine}
	W_{U_T\,\rho\,U_T^\dagger}(a)=W_\rho(T^{-1}(a)).
\end{equation}
Hence, one can unitarily implement the building blocks of Margulis'
random walk.

\section{A quantum Margulis expander}\label{sec:qMargulis}

With these preparations, it is obvious how to proceed. Define the
completely positive map $\Lambda_N$ by
\begin{equation}\label{eqn:quantumMargulis}
	\Lambda_N(\rho) = \frac1{|{\cal S}|} \sum_{T\in {\cal S}} U_T\,\rho\,U_T^\dagger,
\end{equation}
where we have used the notation defined in Eq.\ (\ref{eqn:affine})
above.
One immediately gets:

\begin{observation}[Quantum Margulis expander]
	For odd $N$, 
	the map $\Lambda_N$ (Eq.\ (\ref{eqn:quantumMargulis})) acts on Wigner
	functions in the same way the Margulis expander acts on classical
	probability distributions. In particular, its degree and its
	spectrum are identical to the ones of the Margulis 
	random walk. The
	Wigner functions of $\Lambda$'s eigen-operators are the
	eigen-distributions of the classical random walk.
\end{observation}

{\bf Proof.}
	Let $\Lambda_N^{(C)}$ be the stochastic matrix associated with the
	random walk on the classical Margulis graph. For $v\in\ZZ_N^2$, let
	$e(v)$ be the function on $\ZZ_N^2$, which takes the value 1 at $v$
	and 0 else.
	Clearly, the set $\{e(v)\}_{v\in\ZZ_N^2}$ spans the space of all
	functions on the lattice. Also,
	\begin{equation*}
		\Lambda_N^{(C)}(e(v))=
		\frac1{|\mathcal{S}|}\sum_{T\in\mathcal{S}} e(T(v)).
	\end{equation*}

	Using Eqs.\ (\ref{eqn:affine}, \ref{eqn:quantumMargulis}), we get
	for the quantum version
	\begin{equation*}
		\Lambda_N(A(v))=
		\frac1{|\mathcal{S}|}\sum_{T\in\mathcal{S}} A(T(v)).
	\end{equation*}

	Hence the action of the classical and the quantum expanders are
	identical on a basis. The claims follow.
\square
	
\section{Efficient implementation}

Consider a quantum expander which acts on a tensor-product Hilbert
space $(\CC^d)^{\otimes n}\simeq \CC^N$ for $N=d^n$.  The expander is
\emph{efficient} if it can be realized using
$\operatorname{poly}(n)$ single-qudit or two-qudit quantum gates.
So far, only two efficient constructions have been published
\cite{ben2,harrow}. The Margulis expander adds to this list.

\begin{theorem}[Efficient implementation]\label{thm:efficient}
	The quantized Margulis expander acts efficiently on
	$(\CC^d)^{\otimes n}$.
\end{theorem}

To establish the claim, we need to clarify how we introduce a tensor
product structure in $\CC^N$. Every $0\leq j \leq N-1$ can be
expressed in a $d$-adic expansion as $j=j_1 \dots j_n$ for $0\leq j_l
\leq d$. More precisely, $j=\sum_{l=1}^n j_l d^{n-l}$.  The tensor product
structure is now given by $\ket j=\ket{j_1}\otimes\dots\otimes\ket{j_n}$.

\begin{lemma}[Efficient constituents]
	Let $N=d^n$. The following operators act efficiently on $\CC^N$:
	\begin{enumerate}
		\item
		The quantum Fourier transform
		\begin{equation*}
			F: \ket j \mapsto N^{-1/2} 
			\sum_{k=0}^{N-1} \exp\big({i\frac{2\pi}N \,j k}\big) 
			\ket k.
		\end{equation*}

		\item 
		The Weyl operators $w(1,0)$ and $w(0,1)$. 

		\item
		The operators $\mu(T_1)$ and $\mu(T_2)$.
	\end{enumerate}
\end{lemma}

{\bf Proof. }
	The first statement is well-known. See Chapter 5 in Ref.\
	\cite{nielsen} for the qubit version, which can easily be adapted to
	general $d$.
	Next, consider $w(1,0)=z(1)$. We have
	\begin{eqnarray*}
		z(1)\ket j
		&=& \exp\big(i \frac{2\pi}{d^n}\, j\big) \ket{j_1, \dots, j_n} \\
		&=& \exp\big(i 2\pi\, 
		\sum_{l=1}^n j_l d^{-l}\big) \ket{j_1, \dots, j_n} \\
		&=&
		\bigotimes_l \exp\big(i 2\pi\, j_l d^{-l}\big) 
		\ket{j_l}.
	\end{eqnarray*}
	Hence $z(1)$ is actually local. One confirms that
	$x(1)=F\,z(1)\,F^\dagger$ and thus $x(1)$ is efficient.

	To conclude the proof, we need to borrow three statements from the
	theory of metaplectic representations.
	Firstly, $\mu$ is a projective representation\footnote{Actually,
	$\mu$ is even a \emph{faithful} representation, but that fact is irrelevant
	for our purposes.}, i.e., 
	\begin{equation*}
		\mu(S T)=e^{i \phi(S,T)}\, \mu(S)\mu(T)
	\end{equation*}
	for some phase $\phi(S,T)$
	\cite{marcus,neuhauser,poswig}.
	Secondly,
	\begin{equation*}
		F=
		\mu
		\big(\,
		\left[
		\begin{array}{cc}
			0&1\\
			-1&0
		\end{array}
		\right]
		\,\big),
	\end{equation*}
	and thirdly,
	\begin{equation*}
		U_\pm =
		\mu
		\big(\,
		\left[
		\begin{array}{cc}
			1&\pm 2\\
			0&1
		\end{array}
		\right]
		\,\big)
	\end{equation*}
	is given by
	\begin{eqnarray*}
		U_\pm \ket j 
		= \exp(i2\pi/N\, (\mp j^2))\,\ket j.
	\end{eqnarray*}
	The last two statements can be found in Theorem 4.1 of Ref.\
	\cite{neuhauser} (strictly speaking only for the case of prime $N$,
	but the proofs work for any odd value) or in Lemma 2 to Lemma 4 of
	Ref.\ 
	\cite{marcus}.
	The claim becomes easy to verify:
	\begin{eqnarray*}
		U_\pm \ket j&=& 
		\exp\big(i2\pi\, (\mp 
		\sum_{l,l'=1}^n j_l j_{l'} d^{n-l-l'}
		)\big) 
		\,\ket j\\
		&=&
		\prod_{l,l'} R(l,l') \ket j,
	\end{eqnarray*}
	where we have introduced the diagonal two-qudit unitary
	\begin{equation*}
		R(l,l') \ket{j_l, j_{l'}} = 
		\exp(i2\pi\,  (\mp  j_l j_{l'} d^{n-l-l'}))
		\,\ket{j_l,j_{l'}}.
	\end{equation*}
	Thus $U_\pm$ --  and therefore in particular $\mu(T_1)$ -- 
	are efficient.
	Finally, 
	\begin{equation*}
		T_2=
		\left[
		\begin{array}{cc}
			0&1\\
			-1&0
		\end{array}
		\right]
		\left[
		\begin{array}{cc}
			1&-2\\
			0&1
		\end{array}
		\right]
		\left[
		\begin{array}{cc}
			0&1\\
			-1&0
		\end{array}
		\right]^{-1},
	\end{equation*}
	which implies that $\mu(T_2)\propto F U_- F^{-1}$ is efficient.
\square

The proof of Theorem \ref{thm:efficient} is now immediate, as all the
$U_T$'s which appear in the construction of $\Lambda$ can be
implemented by combining the unitaries treated in the above lemma and
their inverses.

\section{Continuous variable systems}

The quantum phase space terminology of Section \ref{sec:phaseSpace}
has originally been introduced in the context of continuous variable
systems (see e.g.\ Ref.\ \cite{Petz}. In particular, if we
re-interpret the affine transformations $\mathcal{S}$ given in Eq.\
(\ref{eqn:margulis}) as operations on $\RR^2$, we immediately obtain a
completely positive map $\Lambda_\infty$ acting on the
infinite-dimensional Hilbert space of a single mode. Does it
constitute a quantum expander? After reviewing some definitions in
Section \ref{sec:contDefs}, we will give an affirmative answer in
Section \ref{sec:contMarg}. The action of expanders on second moments
is discussed in Section \ref{sec:linialLondon}.

\subsection{Continuous phase space methods}
\label{sec:contDefs}

In the continuous case, the phase space is given by $\RR^2$. Let $X$
and $P$ be the canonical position and momentum operators. 
The  Weyl operators \cite{Petz,Cont,Braun}
are now 
\begin{equation}
	w(p,q)=\exp(i q P - i p X).
\end{equation}
As in Eq.\ (\ref{eqn:parity}), the parity operator $A(0,0)$ acts on state
vectors $\psi\in L^2(\RR)$ as
\begin{equation*}
	(A(0,0)\psi)(x) = \psi(-x).
\end{equation*}
We define the phase space operators $A(p,q)$ for $(p,q)\in\RR^2$ exactly
as in Eq.\ (\ref{eqn:phasespaceoperators}). The Wigner function
becomes
\begin{equation*}
	W_\rho(p,q)=\pi^{-1} \tr\left(A(p,q)\,\rho\right)
\end{equation*}
c.f. Eq.\ (\ref{eqn:wigner}). The obvious
equivalents of Eqs.\ (\ref{eqn:shift},\ref{eqn:meta}) hold for
$a\in\RR^2$ and $S\in \operatorname{Sp}(2,\RR)$,
the group of unit-determinant transformations of the 
two-dimensional real plane.
Hence it is plain how to interpret Eq.\ (\ref{eqn:affine}) and finally
how to turn Eq.~(\ref{eqn:quantumMargulis}) into a definition of
$\Lambda_\infty$, the infinite-dimensional quantum Margulis map.

\subsection{A continuous quantum Margulis expander}
\label{sec:contMarg}

A slight technical problem arises when transferring the definition of
an expander to the infinite-dimensional case: both the invariant
distribution $f(v)=1$ of a classical expander and the invariant
operator $\Id$ of a quantum expander map are not normalizable. Hence,
if we define e.g.\ the action of a completely positive map $\Lambda$
on the set of trace-class operators $\mathcal{T}^1(\H)$, the would-be
eigenvector with eigenvalue 1 is not even in the domain of definition.
In the light of this problem, we switch to the following definition of
a quantum expander, which is compatible with the notion used up to
now.

\begin{definition}
	Let $N\leq \infty$ and set $\H=\CC^N$.
	A completely positive map $\Lambda$ is an
	\emph{$(N,D,\lambda)$-quantum expander} if, for all traceless
	operators $X\in\mathcal{T}^1(\H)$, 
	\begin{equation*}
		||\Lambda(X)||_2 \leq \lambda\,||X||_2.
	\end{equation*}
\end{definition}

The definition above is best understood in terms of the Heisenberg
picture:
\begin{equation*}
	|\tr(\Lambda^n(\rho)\,X)|=|\tr(\rho\,(\Lambda^\dagger)^n(X))|\leq \lambda^n 
\end{equation*}
for all normalized ($||X||_2=1$), traceless observables $X$. Thus the
state becomes ``featureless'' exponentially fast when being acted on
by $\Lambda$.
Let $\lambda_M$ be the second largest eigenvalue of the finite
Margulis expanders. Then:

\begin{observation}[Continuous quantum expander]\label{obs:infinite}
	The infinite-dimensional quantum Margulis map $\Lambda_\infty$ is an
	$(\infty,8,\lambda_M)$-quantum expander.
\end{observation}

Note that by the previous section, we know there are $(N,8,\lambda_M)$
quantum expanders for arbitrarily 
high $N$. A priori, however, this does
not imply the existence of a solution for $N=\infty$.

Once more, by switching to the phase-space picture, the proof of
Observation \ref{obs:infinite} can be formulated completely in
classical terms. The intuition behind the argument is simple to state.
Take an element $T$ of $\mathcal{S}$, e.g.
\begin{equation}
	T: v\mapsto 
	\left[
	\begin{array}{cc}
		1&2\\
		0&1
	\end{array}
	\right]\,v.
\end{equation}
The inverse is given by
\begin{equation}
	T^{-1}=
	\left[
	\begin{array}{cc}
		1&-2\\
		0&1
	\end{array}
	\right],
\end{equation}
regardless of whether the matrix is interpreted as acting on $\RR^2,
\ZZ^2$ or $\ZZ_N^2$. As the same is true for all other elements of
$\mathcal{S}$, the action of the classical Margulis map
``looks similar'' on continuous, infinite discrete and on finite phase
spaces -- at least as long as it acts on distributions which are
concentrated close to the origin, so that the cyclic boundary
conditions of $\ZZ_N^2$ do not come into play. Using this insight, the
following lemma reduces the continuous to the finite case.

\begin{lemma}
	\label{lem:contMargulis}
	Let $f \in C_0^0(\RR^2)$ be a continuous function 
	with compact
	support, such that
	\begin{equation}
		\int_{\RR^2} f(v) dv = 0.
	\end{equation}
	Let $A: L^1(\RR^2)\to L^1(\RR^2)$ be the classical Margulis map
	acting on distributions on $\RR^2$. Then
	\begin{equation}
		||A(f)||_2\leq \lambda_M ||f||_2.
	\end{equation}
\end{lemma}

{\bf Proof. }
	We discretize the problem by partitioning $\RR^2$ into a net of
	squares with side length $\delta$. More specifically, for
	$(x,y)\in\ZZ^2$, let
	\begin{equation*}
		Q_\delta(x,y)=[(x-1/2)\delta,(x+1/2)\delta] \times
		[(y-1/2)\delta,(y+1/2)\delta]
	\end{equation*}
	be the square with edge length $\delta$ centered around $(x\,\delta,
	y\,\delta)\in\RR^2$. The discretized version of $f$ is
	$f_\delta:\ZZ^2\to\CC$ defined by
	\begin{equation*}
		f_\delta(x,y) = \frac1{\delta^2} \int_{Q_\delta(x,y)} f(v)dv.
	\end{equation*}
	Note that $\sum_{x,y}f_\delta(x,y)=0$. 
	On $\ZZ^2$, we use the $\delta$-dependent norm
	\begin{equation*}
		||f_\delta||_2 = \big(\ \delta^2\sum_{x,y}
		|f_\delta(x,y)|^2\big)^{1/2}
	\end{equation*}
	(the factor $\delta^2$ corresponds, of course, to the volume of the
	squares $Q_\delta(x,y)$).
	Now, let $T$ be one of the affine
	transformations in $\mathcal{S}$. We can interpret $T$ as an
	operation on $\ZZ^2$ and define its action on $f_\delta$ accordingly
	by
	\begin{equation*}
		(T(f_\delta))(x,y)=f_\delta(T^{-1}(x,y)).
	\end{equation*}

	For small enough $\delta$, the approximation is going to be
	arbitrarily good: using the uniform continuity of $f$, and the fact
	that all $T\in\mathcal{S}$ are continuous and 
	volume-preserving, one
	finds that for every $\varepsilon>0$, there is a $\delta>0$
	such that simultaneously
	\begin{eqnarray}
		\big|\,||f_\delta||_2 - ||f||_2\,\big| &<&
		\varepsilon/2, \label{eqn:eps1} \\
		\big|\,||A(f_\delta)||_2 - ||A(f)||_2\,\big| &<&
		\varepsilon/2. \label{eqn:eps2}
	\end{eqnarray}

	As the support of $f$ is compact, there is an $R\in\NN$ such that
	$f_\delta(x,y)$ and $A(f_\delta)(x,y)$ are equal to zero 
	whenever
	$|x|\geq R$ or $|y|\geq R$. This 
	enables us to pass from $\ZZ^2$ to
	the finite lattice $\ZZ_{N}^2$ for $N> 2R$. Indeed, when we
	re-interpret $f_\delta$ as a function $\ZZ_{N}^2\to\CC$ and the
	$T\in\mathcal{S}$ as affine transformations on $\ZZ_{N}^2$, the
	values of $||f_\delta||_2$ and $||A(f_\delta)||_2$ 
	remain unchanged.
	But we know that $A$ is an 
	$(N,8,\lambda_M)$-expander for every
	finite $N$. Hence
	\begin{equation*}
		||A(f_\delta)||_2 \leq \lambda_M ||f_\delta||_2,
	\end{equation*}
	implying (by Eqs.\ (\ref{eqn:eps1},\ref{eqn:eps2}))
	\begin{equation*}
		||A(f)||_2 \leq \lambda_M ||f||_2 - \varepsilon.
	\end{equation*}
	This proves the claim, as the right hand side can be chosen to be
	arbitrarily small.
\square

{\bf Proof }\emph{(of Observation \ref{obs:infinite}).}
	Once again, the quantum Margulis map $\Lambda_\infty$ 
	acts on the
	Wigner function $W_X$ of any operator $X$ in the same 
	way the
	classical Margulis scheme acts on distributions on 
	$\RR^2$. Now,
	$X\in\mathcal{T}^1(\H)$ implies $W_X\in L^2(\RR^2)$. Because
	$C_0^0(\RR^2)$ is dense in $L^2(\RR^2)$ and $\Lambda_\infty$
	is continuous, Lemma \ref{lem:contMargulis}
	suffices to establish the claim.
\square

\subsection{Action on second moments}
\label{sec:linialLondon}

In physics, one often measures the concentration of a phase space
distribution by its second moments with respect to canonical
coordinates. Thus, it may be interesting to look for signatures of the
strong mixing properties of a quantum expander in its action on second
moments. 

More precisely, first moments are the expectation values of the
position and momentum operators $(\langle X\rangle,\langle
P\rangle)^T$ (where $\langle A \rangle = \tr(\rho A)$ for an operator
$A$).  The second moments are defined as the entries of the
\emph{covariance matrix}:
\begin{equation*}
	\gamma=2\operatorname{Re}\left[
	\begin{array}{cc}
	  \langle X^2 \rangle -  \langle X \rangle^2& \langle XP  \rangle -
		\langle X \rangle\langle P \rangle\\ \langle PX \rangle - \langle X \rangle\langle P \rangle &   \langle P^2 \rangle
	-   \langle P \rangle^2
	\end{array}
	\right].
\end{equation*}
As the action of the continuous quantum expander in state space is
defined via the metaplectic representation, the change in second
moments can be computed explicitly. 
In particular, any $S\in \operatorname{Sp}(2,\RR)$ gives rise to a
congruence $\gamma\mapsto S\gamma S^T$ for second moments. More generally, it is not difficult to see that for arbitrary 
convex combinations of states subject to
affine transformations, the output's first and second moments depend
only on the same moments of the input.

Under the Margulis random walk, one obtains for the first moments
\begin{eqnarray*}
	\langle X\rangle\mapsto \frac{1}{|{\cal S}|}
	\sum_{T\in {\cal S}} x_T,\quad
	\langle P\rangle\mapsto 
	\frac{1}{|{\cal S}|} 
	\sum_{T\in {\cal S}} p_T
\end{eqnarray*}
with $(x_T,p_T)^T = T(\langle X\rangle, \langle P\rangle)^T$. For the
second moments:
\begin{eqnarray}\label{CVE}
	\gamma
	\mapsto f(\gamma):=
	\sum_{i=1}^2 \Big(
	T_i \gamma T_i^T+T_i^{-1} \gamma (T_i^{-1})^T
	\Big)+2\, G,
\end{eqnarray}
where the matrix $G$ is given by
\begin{equation*}
	G=\left[
	\begin{array}{cc}
	\sum_T  \frac{x_T^2}{|{\cal S}|} - (\sum_T 
	\frac{x_T}{|{\cal S}|} )^2 &
	\sum_{T} 
	\frac{x_T p_T}{|{\cal S}|} - \sum_{T,T'}  
	\frac{x_T p_{T'}}{|{\cal S}|^2},	\\
	\sum_{T} 
	\frac{x_T p_T}{|{\cal S}|} - \sum_{T,T'}  
	\frac{x_T p_{T'}}{|{\cal S}|^2}	
	& \sum_T  \frac{p_T^2}{|{\cal S}|} - (\sum_T 
	\frac{p_T}{|{\cal S}|} )^2
	\end{array}
	\right].
\end{equation*}
The latter matrix is evidently positive 
\cite{Positive}. To show that the main diagonal
entries of $f^{(n)}(\gamma)$
diverge exponentially in the number 
$n$ of applications of the map $f$, 
it is hence sufficient to consider the map
\begin{eqnarray*}
	\gamma
	\mapsto g(\gamma) = 
	\sum_{i=1}^2 \Big(
	T_i \gamma T_i^T+T_i^{-1} \gamma (T_i^{-1})^T
	\Big),
\end{eqnarray*}
since
\begin{equation*}
	f^{(n)}(\gamma)\geq g^{(n)}(\gamma).
\end{equation*}	
%
A simple calculation yields 
\begin{equation*}
	\gamma= \left[
	\begin{array}{cc}
	a & b\\
	b & c
	\end{array}
	\right] 
	\mapsto g(\gamma)= 
	\left[
	\begin{array}{cc}
	a+2c & b\\
	b & c+2a
	\end{array}
	\right].
\end{equation*}
Let $\gamma^{(n)}= g^{(n)}(\gamma)$ 
be the covariance matrix after $n$ 
iterations of $g$
and define 
$\alpha=(a+c)/2$, and $\beta = (a-c)/2$ to simplify notation.
Then
\begin{eqnarray*}
	\gamma^{(n)}
	&=&
	\left[
		\begin{array}{cc}
			3^n \alpha  + (-1)^n \beta & b\\
			b & 3^n \alpha - (-1)^n \beta 
		\end{array}
	\right] .
\end{eqnarray*}
This means that 
\begin{eqnarray*}
	\frac1n \log_3 (\gamma^{(n)})
	&\to&
	\left[
		\begin{array}{cc}
			1 & 0 \\
			0 & 1
		\end{array}
	\right]\qquad (n\to \infty).
\end{eqnarray*}
Thus, the elements of the main diagonal -- and therefore
also 
$\tr(f^{(n)}(\gamma) ), \det( f^{(n)}(\gamma) )$, and
$\operatorname{spec}( f^{(n)}(\gamma) )$ -- diverge 
exponentially in the number $n$ of iterations.

\section{Summary and Outlook}

Employing phase space methods, we were able to quantize a
well-established combinatorial structure with almost no technical
effort. Until now, discrete Wigner functions have been studied mainly
for their mathematical appeal. As far as we know, the present work is
the first instance where a problem not related to the phase space
formalism itself has been solved using the properties of discrete
Wigner functions.

The unitaries which appear in the construction of expanders have
randomization properties which are in some sense 
extremal. It would be
interesting to see whether connections to other extremal sets of
unitaries -- e.g., {\it unitary designs} \cite{U1,U2} -- can be
found.  Also, more practical applications may be anticipated, e.g.,
when one aims at initializing quantum systems in the maximally
mixed state with few (i.e. $D$) operations, under repeated 
invocation of the same completely positive map $\Lambda$. 
From a physics perspective, the existence of quantum 
expanders gives rise to the quite counterintuitive 
observation that small environments can lead to
strongly mixing and decohering dynamics: One may think
of a controlled-$U$ operation, the unitaries being those
from a quantum expander, where the control has 
the dimension of a mere $\log_2(D)$. Lastly,
the programme may improve the understanding of 
iterated {\it randomization procedures}, as the one discussed
in Ref.\ \cite{toth}.

{\it Acknowledgments:} We thank T.J.\ Osborne and
A.\ Harrow for discussions on quantum expanders. DG is pleased to
acknowledge the hospitality of M.\ M\"uller of the Mathematical
Physics Group at the TU Berlin, where this work has been initiated.
Supported has been provided by the 
EU (QAP, COMPAS), 
the QIP-IRC, Microsoft Research, 
and the EURYI Award Scheme.


\begin{thebibliography}{99}

\bibitem{survey}
	S.\ Hoory, N.\ Linial, and A. Widgerson,
	Bulletin of the American Mathematical 
	Society {\bf 43}, 439 (2006).
	
\bibitem{hastings1}
	M.B.\ Hastings, Phys.\ Rev.\ B {\bf 76}, 035114 (2007).

\bibitem{hastings2}
	M.B.\ Hastings, Phys.\ Rev.\ A {\bf 76}, 032315 (2007).

\bibitem{ben1}
	A.\ Ben-Aroya and A.\ Ta-Shma, 
	arXiv:quant-ph/0702129 (2007).

\bibitem{ben2}
	A.\ Ben-Aroya, O.\ Schwartz, and
	A.\ Ta-Shma, arXiv:0709.0911 (2007).

\bibitem{harrow}
	A.\ Harrow, Quant.\ Inf.\ Comp.\
	{\bf 8}, 0715 (2008).

\bibitem{ambainis}
	A.\ Ambainis and A.\ Smith, 
	Proceedings of RANDOM 2004, Cambridge,
	MA (2004), quant-ph/0404075; P.\ Dickinson and A.\ Nayak,
	Proceedings of Quantum Computing Back Action, Kanpur (2006)
	quant-ph/0611033;
	I.\ Kerenidis and D.\ Nagaj, 
	J.\ Math.\ Phys.\ {\bf 47}, 092102 (2006).

\bibitem{Mixed}
	It follows directly from the definition of an
	$(N,D,\lambda)$-quantum
	expander that 
	\begin{equation}
		\| \Lambda_N(\rho) - \Id/N\|_2\leq \lambda
		 \| \rho -  \Id/N\|_2,
	\end{equation}
	that is, each invocation of the quantum expander
	contracts the $2$-norm distance to the maximally
	mixed state in $\CC^N$ by at least $\lambda$.
	
\bibitem{margulis}
	G.\ Margulis, Problemy Peredaci Informacii, {\bf 9}, 
	71 (1973).

\bibitem{gabber}
	O.\ Gabber and Z.\ Galil, J.\ Comput.\ System Sci.\ 
	{\bf 22}, 407
	(1981).

\bibitem{hudson}
	R.L.\ Hudson, 
	Rep.\ Math.\ Phys.\ {\bf 6}, 249 (1974).

\bibitem{poswig}
	D.\ Gross, 
	J.\ Math.\ Phys.\ {\bf 47}, 122107 (2006).

\bibitem{poswigShort}
	D.\ Gross, 
	Appl.\ Phys.\ B {\bf 86}, 367 (2007).

\bibitem{gottesman}
		D.\ Gottesman, 
		{\it Stabilizer codes and quantum error correction}.
		PhD thesis, Caltech (1997). quant-ph/9705052.

\bibitem{neuhauser}
		M.\ Neuhauser,
		Journal of Lie Theory {\bf 12}, 15 (2002).

\bibitem{vourdas}
		A.\ Vourdas, Rep. Prog. Phys. {\bf 67}, 267 (2004).

\bibitem{marcus}
		D.\ Appleby, J.\ Math.\ Phys.\ {\bf 46}, 052107 (2005).

\bibitem{nielsen}
	M.A.\ Nielsen and I.L.\ Chuang,
	{\it Quantum computation and quantum information}
	(Cambridge Univ. Press, Cambridge, 2000).

\bibitem{Petz}	
	D.\ Petz,  {\it An invitation to the algebra of the 
	canonical commutation relation} 
	(Leuven University Press, Leuven, 1990).

\bibitem{Cont}
	J.\ Eisert and M.B.\ Plenio,
	Int.\ J.\ Quant.\ Inf.\ {\bf 1}, 479 (2003).
	
\bibitem{Braun}
	S.L.\ Braunstein and P.\ van Loock, 
	Rev.\ Mod.\ Phys.\ {\bf 77}, 513 (2005).

\bibitem{Positive}
	This can be shown by writing $G$ as
	$G=A A^T$ with
	$A\in \RR^{2,|{\cal S}|}$ with entries
	\begin{eqnarray}
		A_{1,T}&=& \frac{x_T}{|{\cal S}|^{1/2}}- \sum_{T'} 
		\frac{x_T'}{|{\cal S}|^{1/2}},\\
		A_{2,T}&=& \frac{p_T}{|{\cal S}|^{1/2}}- \sum_{T'} 
		\frac{p_T'}{|{\cal S}|^{1/2}}.
	\end{eqnarray} 
		
\bibitem{U1}
	C.\ Dankert, R.\ Cleve, J.\ Emerson, and E.\ Livine, 
	quant-ph/0606161.
	
\bibitem{U2}
	D.\ Gross, K.\ Audenaert, and J.\ Eisert,
	J.\ Math.\ Phys.\ {\bf 48}, 052104 (2007).	
	
\bibitem{toth}
	G.\ Toth and J.\ Garcia-Ripoll,
	Phys. Rev. A {\bf 75}, 042311 (2007).

\end{thebibliography}
\end{document}